\documentclass[aps,prb,reprint,superscriptaddress,floatfix,bibnotes]{revtex4-1}
\usepackage{epstopdf}
\usepackage{graphicx} 
\usepackage{amsmath,amssymb}

\begin{document}

\title{Stratified Graphene-Noble Metal Systems for Low-Loss Plasmonics Applications}

\author{L.\ Rast}
\email[Electronic mail: ]{annalauren.rast@nist.gov}
\affiliation{Applied Chemicals and Materials Division, National Institute of Standards and Technology, Boulder, CO 80305}

\author{T.\ J.\ Sullivan}
\affiliation{Mathematics Institute, University of Warwick, Coventry, CV4 7AL, United Kingdom}
\author{V.\ K.\ Tewary}
\affiliation{Applied Chemicals and Materials Division, National Institute of Standards and Technology, Boulder, CO 80305}


\begin{abstract}
We propose a composite layered structure for tunable, low-loss plasmon resonances, which consists of a noble-metal thin film coated in graphene and supported on a hexagonal boron nitride (hBN) substrate. We calculate electron energy loss spectra (EELS) for these structures, and numerically demonstrate that bulk plasmon losses in noble-metal films can be significantly reduced, and surface coupling enhanced, through the addition of a graphene coating and the wide-bandgap hBN substrate. Silver films with a trilayer graphene coating and hBN substrate demonstrated surface plasmon-dominant spectral profiles for metallic layers as thick as $34\,$nm. A continued-fraction expression for the effective dielectric function, based on a specular reflection model which includes boundary interactions, is used to systematically demonstrate plasmon peak tunability for a variety of configurations. Variations include substrate, plasmonic metal, and individual layer thickness for each material. Mesoscale calculation of EELS is performed with individual layer dielectric functions as input to the effective dielectric function calculation, from which the loss spectra are directly determined. 
\end{abstract}
\maketitle

\section{Introduction}
Plasmonic devices hold promise for a host of metamaterials applications, due to their ability to control light propagation on a subwavelength scale. Noble metal nanomaterials are in some ways ideal components in such devices -- they possess tunable, large amplitude plasmon resonances, which can be excited at optical wavelengths.~\cite{Johnson1972} Nevertheless, metallic plasmonic devices at infrared and visible wavelengths present significant challenges due to bulk plasmon losses. These large losses severely limit the practicality of these materials for a wide variety of applications, particularly in telecommunications and photovoltaics.~\cite{west2010} 

Bulk losses may be mitigated through the use of very thin noble-metal films; however, fabrication of uniform thin metal films is experimentally challenging. Thin films tend to form islands and often require adhesion layers that alter the electronic structure of the device.~\cite{Boragno2009,Rha1997} Additionally, thin noble-metal films still suffer significant resistive losses in the visible regime.~\cite{Atwater2010} Composite materials designed to shift plasmon resonances to a low-loss regime, while increasing coupling to surface plasmons and diminishing bulk resonances, provide clear advantages from a fabrication standpoint. Systems that combine the strong and tunable optical-wavelength plasmon resonances of noble metals with materials possessing improved transport properties could mitigate surface-plasmon losses in the visible regime. 

Graphene is an ideal candidate for such a composite structure due to its unparalleled carrier mobility. This allows for extremely enhanced and tunable electromagnetic response spectra when doped with other plasmonic materials, or fabricated as a component of a multilayer structure. Monolayer graphene has a response spectrum dominated by absorption peaks at $\approx 4.5\,$eV and $\approx 15\,$eV, the $\pi$ and $\pi$ + $\sigma$ surface plasmons. Graphene plasmon resonance provides low losses in the frequency regime below the optical phonon frequency of $0.2\,$eV, where large losses are typically present for metallic plasmonic materials.~\cite{Geim2007} 

Hexagonal boron nitride (hBN) provides several advantages as a substrate for graphene-based plasmonic materials, as it is an ultra-flat wide bandgap insulator with excellent thermal transport properties.~\cite{Xue2011} The flat morphology and uniformity provided by hBN are are desirable as key factors in preserving the transport properties of graphene.~\cite{Xue2011} Results of our numerical study clearly elucidate the advantages of the hBN spectral response for low-loss plasmonics applications. We demonstrate in this work that through the combined advantages of graphene and noble-metal films, along with careful choice of substrate material and layer thickness, multilayer systems can be tuned for low-loss surface plasmon resonance (SPR). 

Metamaterials derive their exotic properties in part from advantageous behaviors of their constituent materials, as well as collective behaviors that emerge due to interactions within the system. The complex task of designing metamaterials tailored for applications as diverse as photovoltaics, biosensing, and microscopy gives rise to a need to develop efficient methods for predicting metamaterial properties. These methods must realistically treat both the individual material properties and the interactions among the constituent materials. We detail and employ such a method for the calculation of electron energy loss spectra (EELS) of multilayer structures consisting of graphene layers on noble-metal (silver, gold, and copper) films with silicon and hBN substrates. 

The effective dielectric function is based on a specular reflection model, first derived by Lambin et al.~\cite{Lambin1985}, and takes into account the boundary conditions across each layer in the stratified structure. The use of the efficient continued-fraction expression along with pre-prepared libraries of dielectric functions for the individual materials opens up the possibility of multilayer graphene composites by design. 

\section{EELS Calculation Details}
\subsection{General Procedure}
\begin{figure}
\includegraphics[scale=1.0]{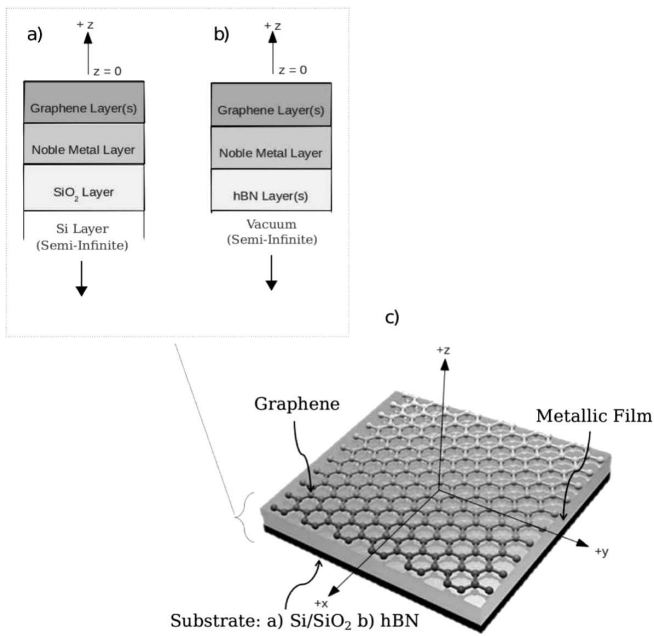}
\caption{\label{Multilayer}Multilayer structure: EELS are calculated for a structure consisting of graphene top layer(s), noble-metal middle layer, and semiconductor substrate.}
\end{figure}
EELS are calculated for a variety of multilayer sandwich structures as depicted in Fig.~\ref{Multilayer}. In general, the sandwich structure consists of a noble-metal middle layer, a graphene coating of an integer number of layers, and a semiconducting substrate. Graphene layer numbers vary from 1 to 20. Noble metals employed include silver, gold, and copper, and are varied in thickness. The effect of two different substrates is considered: (1) a more traditional substrate, Si with a $4\,$nm top layer of SiO$_2$ (Fig.~\ref{Multilayer} (a)), and (2) hBN (Fig.~\ref{Multilayer} (b)). $4\,$nm is a typical thickness for the SiO$_2$ layer, which arises from the thermal processing of the Si substrate.~\cite{Seah2002} The Si layer is assumed to be semi-infinite. In the case of the hBN substrate, an integer number of hBN layers is considered, with a semi-infinite vacuum layer below the hBN layer. The single-layer thickness for both graphene and hBN are taken to be $3.4\,$\AA, the natural inter-layer spacing for graphite.~\cite{arxiv2011} In all cases, a semi-infinite vacuum layer exists above the top ($z = 0$) layer. 

Individual complex dielectric functions are obtained for each layer. We calculate the values through ab-initio methods for graphene and hBN. Empirical values from the literature are used for the metal and silicon substrate layers. These values are then stored for use as input to a continued-fraction algorithm, which yields the effective dielectric function. This algorithm is outlined in Section~\ref{composite}.

\subsection{\label{composite}The Effective Dielectric Function}
The effective dielectric function, $\xi(\omega,k,z)$, of the stratified structure in Fig.~\ref{Multilayer} is that of Lambin et al.~\cite{Lambin1985} The expression for $\xi$ was derived from EELS theory in a reflection geometry. The expression has been shown to be applicable to both phonons~\cite{Lambin1985} and polaritons~\cite{PhysRevB.38.5438} in stratified structures with histogram-like dielectric functions (continuous within each layer) and interacting interfaces. Though Lambin's initial paper containing the derivation focused on semiconducting materials, the expression and the formalism from which it is derived also apply to surface plasmon structure of alternating of metal-insulator layers.~\cite{PhysRevB.38.5438,PhysRev.182.539} It is worth noting that the specular reflection model agrees well with the spectroscopic predictions of the Bloch hydrodynamic model in the small wave vector regime considered in our work.~\cite{Ritchie1966234}

The $z$ coordinate is in the direction perpendicular to the free surface of the sample, extending from the $z = 0$ surface to $-\infty$. $\mathbf{k}$ denotes the surface excitation (plasmon or phonon) wave vector and $\omega$ is the frequency of excitation.
 
\begin{equation}
\xi (\mathbf{k},\omega,z) = \frac{i\mathbf{D}(\mathbf{k},\omega,z)\cdot \mathbf{n}}{\mathbf{E}(\mathbf{k},\omega,z)\cdot\mathbf{k}/k},
\end{equation}
where $\mathbf{D}(\mathbf{k},\omega,z) = \epsilon(\omega,z)\mathbf{E}(\mathbf{k},\omega,z)$, and $\epsilon(\omega,z$) is the long wavelength dielectric function (tensor) of the material at $z$. $\xi$ remains continuous even in the case of sharp interfaces parallel to the $x$-$y$ directions below the surface (as is the case in our multilayer system). This is due to the interface boundary conditions: continuity of $D_\perp$ and $E_\parallel$. 

The effective dielectric function $\xi_0(k,\omega)$ (Eq.~\ref{continuedfrac}) is a solution to the Riccati equation (Eq.~\ref{ricatti}), in the long-wavelength approximation $k\approx0$, at the $z=0$ surface.~\cite{Lambin1985} We fix $k$ as $k = 0.05\,$\AA$^{-1}$ for both the ab-initio calculations and the composite calculation. 
Eq.~\ref{ricatti} was derived for heterogeneous materials made of a succession of layers (with homogeneous dielectric functions within each layer), the layers having parallel interfaces. $\epsilon(z)$ are complex functions, with positive imaginary parts at $z=0$.~\cite{Lambin1985} 

\begin{eqnarray}\label{ricatti}
\frac{1}{k}\frac{\mathrm{d}\xi(z)}{\mathrm{d}z} + \frac{\xi^2(z)}{\epsilon(z)} = \epsilon(z)
\end{eqnarray}
\begin{equation}\label{continuedfrac}
\xi_0 = a_1 - \frac{b_1^2}{a_1+a_2-\frac{b_2^2}{a_2+a_3-\frac{b_3^2}{a_3+a_4-\cdots }}}
\end{equation}
where 
\begin{eqnarray}
a_i=\epsilon_i\coth(kd_i)
\end{eqnarray}
and 
\begin{eqnarray}
b_i=\epsilon_i / \sinh(kd_i).
\end{eqnarray}

Once individual dielectric functions are obtained, this procedure allows for the performance of mesoscale EELS calculations of a wide variety of structures. Layer thickness and materials can easily be substituted in the calculation, with each EELS calculation running in a fraction of a second (nearly independent of the spectral range). EELS are calculated directly from the effective dielectric function as
\begin{equation}\label{EELS}
\mathrm{EELS} = \mathrm{Im}\left[\frac{-1}{\xi(\omega,k) + 1}\right].
\end{equation}

Inspection of Eq.~\ref{continuedfrac} reveals that for $\mathrm{Im}[\epsilon_i] > 0$, $\mathrm{Im}[\xi_0]>0$. The EELS spectra given by Eq.~\ref{EELS} are then positive.

\subsection{Noble-Metal Dielectric Functions}
The copper, silver, and gold complex dielectric functions are empirical values by Johnson and Christy~\cite{Johnson1972} obtained by reflection and transmission spectroscopy on vacuum-evaporated films at room temperature. Film-thickness in the Johnson and Christy study ranged from $185\,$\AA -- $500\,$\AA. Dielectric functions in the film-thickness range $250\,$\AA\ -- $500\,$\AA\ did not vary significantly. In our work, the intermediate value $340\,$\AA\ was chosen to represent bulk mode dominant (yet still nanoscale) metallic thin films. 
\subsection{SiO$_2$ and Si Dielectric Constants}
Relative static permittivities of 3.9 and 11.68 are used for the SiO$_2$ and Si dielectric constants, respectively. These are reasonable and widely-used values from the literature.~\cite{muraka2003, yi2012}
\subsection{Graphene and hBN Individual Layer Dielectric Functions}
\begin{figure}
\includegraphics[scale=1.0]{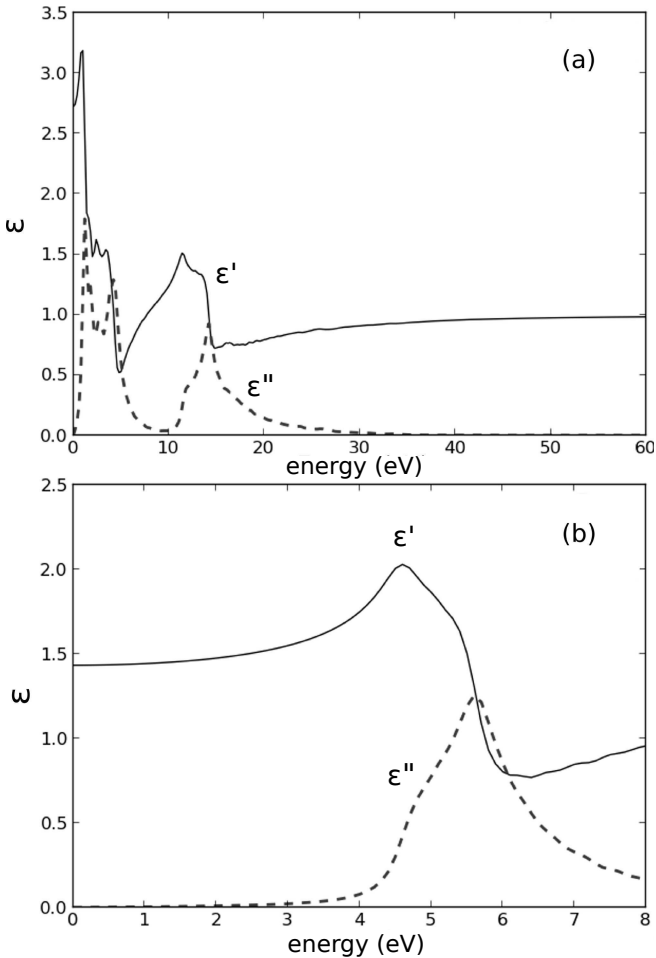}
\caption{\label{grandhbn} Complex relative dielectric function $\epsilon(\omega)$ for graphene (a) and hBN (b). Real and imaginary parts ($\epsilon'(\omega)$ and $\epsilon''(\omega)$) are represented by solid and dotted lines, respectively.}
\end{figure}

Complex dielectric functions for graphene and hBN are displayed in Fig.~\ref{grandhbn} (a) and Fig.~\ref{grandhbn} (b), respectively. These ab-initio calculations use the time-dependent density functional
theory in the local density approximation (LDA), and are implemented in the Python code GPAW, a real-space electronic structure code using the projector augmented wave method.~\footnote{Certain commercial equipment, instruments, or materials are identified in this paper in order to specify the experimental procedure adequately. Such identification is not intended to imply recommendation or endorsement by the National Institute of Standards and Technology, nor is it intended to imply that the materials or equipment identified are necessarily the best available for the purpose.}$^{,}$~\cite{gpaw1, gpaw2, gpaw3, gpaw4} Both graphene and hBN dielectric functions are calculated in the armchair configuration for a momentum transfer value of $0.05\,$\AA$^{-1}$, along the $\bar{\Gamma}$-$\bar{M}$ direction of the surface Brillouin zone. The armchair configuration and value of $k$ are selected for comparison of EELS with existing data in the literature obtained via density functional theory (DFT) methods. The $k$-point sampling with $30 \times 30 \times 1$ Monkhorst--Pack grid was chosen for the band-structure and EELS calculations for both graphene and hBN.

Our model utilizes dielectric functions due to surface parallel excitations only, as the effective dielectric function is derived in a specular reflection geometry. This is a reasonable approximation as out-of-plane excitations are minimal in graphene at energies less than $\approx 10\,$eV, and extreme UV radiation is outside of the regime of interest for this study.~\cite{Gass2008} A lattice constant of $2.46\,$\AA\ is used for graphene, and $2.50\,$\AA\ for hBN, as hBN is nearly isomorphic to the graphene, except for the slightly larger lattice constant.~\cite{Song2010} 

The dielectric function we have obtained for graphene (see Fig.~\ref{grandhbn} (a)) is nearly identical to those of Yan et al.~\cite{gpaw4}, where the authors used the projector augmented wave methodology implemented in GPAW  in terms of linear combinations of atomic orbitals, with a momentum transfer of $0.046\,$\AA$^{-1}$, along the $\bar{\Gamma}$-$\bar{M}$ direction of the surface Brillouin zone. Both our results and that of Yan et al.\ display collective peaks for free-standing single-layer graphene at  $\approx 5\,$ eV and  $\approx 15\,$ eV.  These values also agree with experimental EELS results for single-layer graphene plasmons (with in-plane excitation), for example Eberlein et al.~\cite{PhysRevB.77.233406} find the $\pi$ plasmon at 4.7 eV and $\pi$ + $\sigma$ at 14.6 eV.

A low-energy peak (below 1 eV) is also apparent in Fig.~\ref{grandhbn} (a).  This feature, which has been observed in ab-intio calculations by others~\cite{PhysRevLett.101.226405,gpaw4}, corresponds to a broad shoulder in the EELS, and is due to the low-energy $\pi$ $\rightarrow$ $\pi^{\star}$ single-particle excitation. Low energy graphene features, including intraband transitions are generally quite dependent upon the value of momentum transfer.~\cite{PhysRevB.75.205418} These features are outside the energy regime of interest for the surface and bulk plasmon peaks in this paper, work in which very low-energy excitations are a focus should not neglect momentum transfer dependence. In the context of these calculations, increasing momentum transfer values lead to a shift towards higher energies of the $\pi$ plasmons for intrinsic graphene.~\cite{gpaw4,Gao20111009} Others, for example Gao et al. ~\cite{Gao20111009} have found (via the time-dependent local density approximation) linear dispersion for the $\pi$ plasmon (the key feature in our energy regime of interest) in single-layer graphene for both $\bar{\Gamma}$-$\bar{M}$ and $\bar{\Gamma}$-$\bar{K}$ directions. The calculations of Gao et al. were found to agree well with experiment.

Our hBN dielectric function (see Fig.~\ref{grandhbn} (b)) compares well with that of Yan et al.~\cite{PhysRevB.86.045208}, who obtained their results in the long-wavelength limit, using the LDA-adiabatic local density approximation method. Both our results for hBN spectra and that of Yan et al. display a broad absorbtion peak with onset at $\approx 4.5\,$ eV and maximum at $\approx 5.75\,$ eV, which is in good agreement with previous literature.~\cite{PhysRevLett.96.126104} The dielectric function obtained for hBN is also similar to the experimental results of Tarrio and Schnatterly~\cite{Tarrio}, where the authors measured a peak in the imaginary part of the dielectric function at $\approx 6\,$ eV. Differences between the complex dielectric function used in this study and that of ~\cite{Tarrio} are attributable to the larger momentum transfer value of $0.13\,$\AA$^{-1}$ in the Tarrio and Schnatterly study. 

\section{RESULTS}
\subsection{Decreasing metallic film thickness and the begrenzung effect}
\begin{figure}
\includegraphics[scale=1.0]{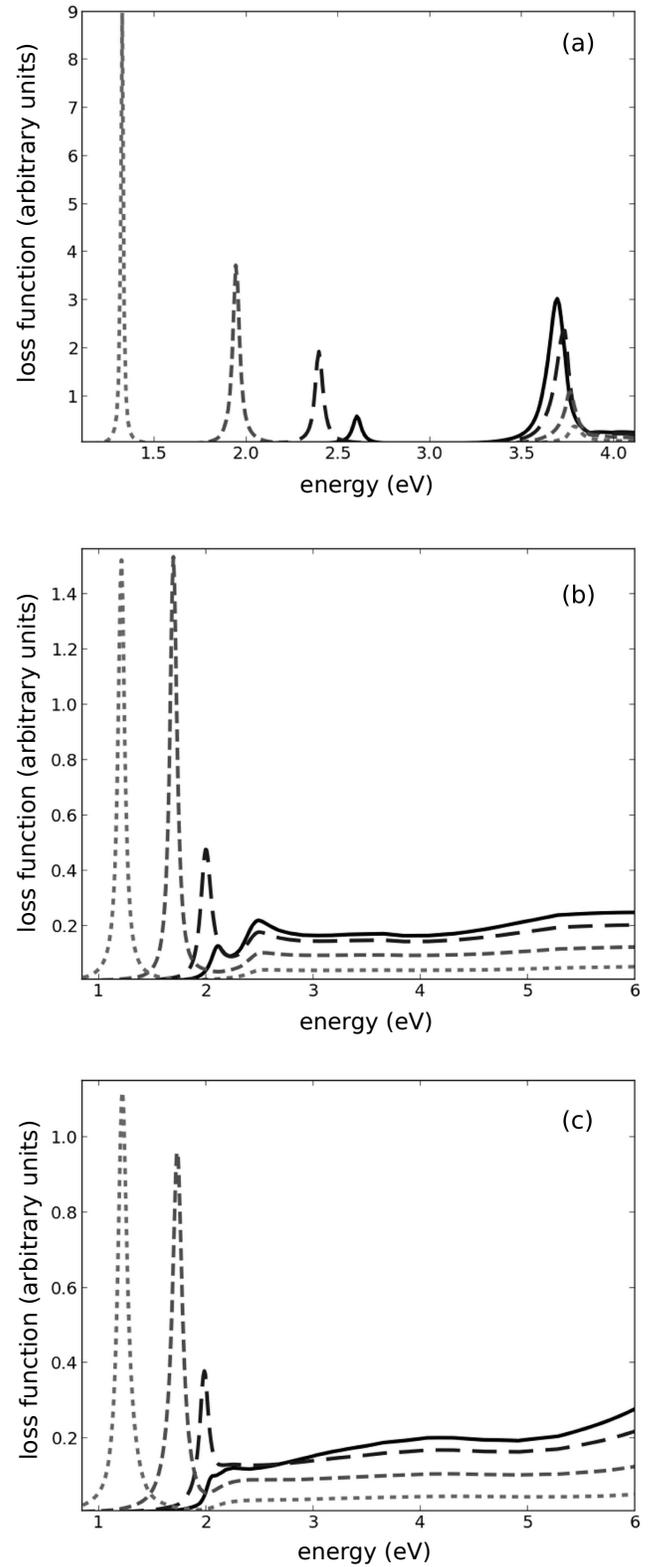}
\caption{\label{Sinogrsilverthicknesses}EELS: (a), (b), and (c) demonstrate the effect of differing thickness for the Ag, Au, and noble-metal layer, respectively, with a SiO$_2$/Si substrate, and without a graphene coating. Film thicknesses are $34\,$nm (solid line), $20\,$nm (long dashes), $10\,$nm (short dashes), and $4\,$nm (dotted line).}
\end{figure}
Figures ~\ref{Sinogrsilverthicknesses} (a)--(c) (respectively, Ag, Au, and Cu on SiO$_2$/Si, without graphene top coating) demonstrate the effect of decreasing noble metal layer thickness. These data should serve as a basis for comparison with Fig.~\ref{silvergr2} (a)--(c), which are discussed in detail in the next section. 

As the noble metal film thickness is reduced in Fig.~\ref{Sinogrsilverthicknesses} (a), the so called \emph{begrenzung effect} is apparent. An increase in the surface-to-volume ratio in the metal causes enhanced coupling to the surface resonance and diminished coupling to the bulk modes.~\cite{Ritchie1957, Osma1997} 

In the case of a thin metallic slab, empirical models have been quite thoroughly explored. Upon the introduction of a boundary to an infinite metallic slab, a negative (begrenzung) peak is introduced at the same energy as the bulk peak, and a trailing surface peak appears.~\cite{Ritchie1957} The surface peak becomes more pronounced with decreasing thickness, as does the negative begrenzung peak, decreasing the net bulk-plasmon amplitude. A sharp transition can be observed, between $34\,$nm and $10\,$nm Ag film thickness, in which surface modes become dominant. This transition can also be observed for Au and Cu films, between $34\,$nm and $20\,$nm. This is consistent with observations in the well-validated and widely-used empirical data by Johnson and Christy.~\cite{Johnson1972}

\subsection{SPR Enhancement due to the graphene coating}
\subsubsection{SiO$_2$/Si Substrate}
\begin{figure}
\includegraphics[scale=1.0]{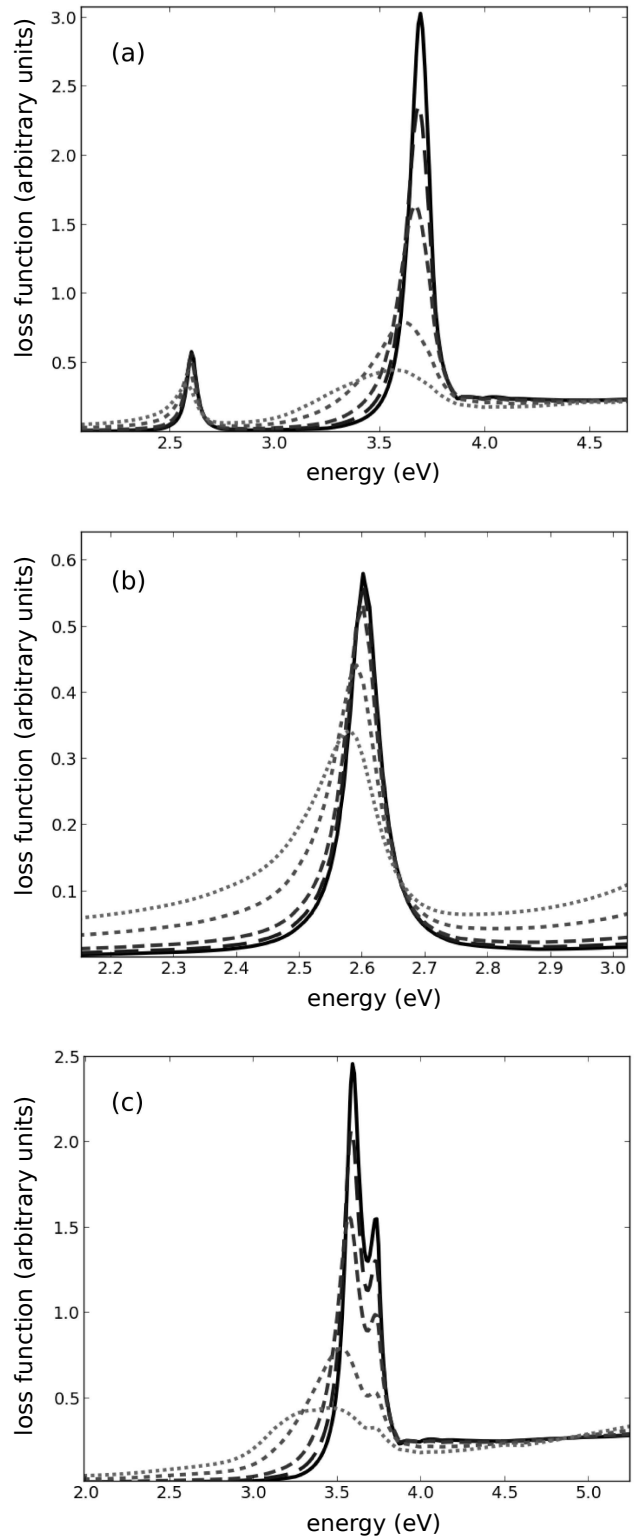}
\caption{\label{silvergr2}EELS for varying numbers of layers for the graphene film: (a) $34\,$nm Ag layer and SiO$_2$/Si substrate, (b) the surface peak in (a), and (c) $34\,$nm Ag layer and monolayer hBN substrate. In both cases the graphene layer numbers are 0 (solid line), 1 (long dashes), 3 (intermediate-length dashes), 10 (short dashes), and 20 (dotted line).}
\end{figure}
Figures~\ref{silvergr2} (a) and (b) demonstrate the effect of the addition of graphene coatings of varyious thicknesses to the Ag surface, in the case of a SiO$_2$/Si substrate. For up to three layers of graphene, the surface peak remains virtually unchanged. However, at three layers, the bulk peak amplitude is reduced by approximately $50 \%$. This effect is attributed to the introduction of a thin boundary layer, decreasing the bulk resonance amplitude. Imposition of a boundary leads to simultaneous diminished coupling to bulk modes and enhanced coupling to surface modes, physically similar to the aforementioned begrenzung effect.~\cite{Ritchie1957, Osma1997, egerton2011} At 20 layers, as we would expect, the bulk peak broadens significantly (indicating increased losses) and the system has spectral properties (broadening and peak position) resembling those of  graphite on silver. Peak positions calculated for the 20-layer graphene coating ($\approx 2.6\,$eV and $\approx 3.5\,$eV) are very close to those measured in EELS of thin films of Ag nanoparticles evaporated on graphitic surfaces, for example $2.2\,$eV and $3.4\,$eV.~\cite{Palmer2002} Differences in peak position (particularly for the lower-energy peak) are attributable to differences in morphology between the Ag slab in our calculations and the film of Ag nanoparticles in the experimental samples. 

\subsubsection{hBN Substrate}
Figure~\ref{silvergr2} (c) demonstrates the effect of the addition of graphene coatings of varying thickness to the silver surface, in the case of a hBN substrate. The use of the hBN substrate dramatically enhances the SPR peak, as well as shifting the surface peak to an energy very close to that of the bulk peak, even the case where no graphene coating is used. This blue-shift of the SPR is due to the much smaller real part of the hBN dielectric function in comparison to Si. Decreasing substrate dielectric function is known to dramatically blue-shift SPR.~\cite{dielectrictuning}

Addition of a single-layer graphene coating serves to further diminish the bulk peak without significant degradation of the surface peak. For up to 3 layers of graphene, the surface peak does not broaden significantly. As in the case of the SiO$_2$/Si substrate, broadening of the surface peak for more than 3 layers of graphene corresponds to increased losses. Indeed, at 10 graphene layers the response is what one would expect from a graphite coating. A similar transition in plasmonic behavior at 10 layers has been observed experimentally for multilayer graphene EELS.~\cite{PhysRevB.77.233406}
\subsubsection{Ideal configurations for Ag: enhancement of surface modes}
\begin{figure}
\includegraphics[scale=1.0]{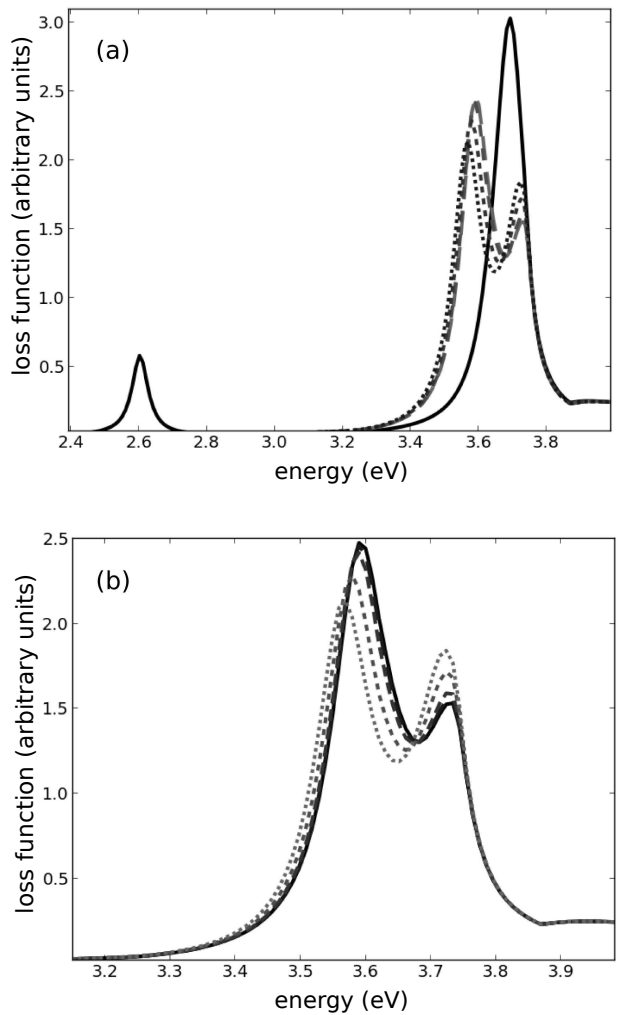}
\caption{\label{monolayerhbnvseverything.png}EELS:(a) Comparison of the use of a SiO$_2$/Si substrate (solid line), a mono-, tri-, 10-, and 20-layer hBN substrate (long dashes, intermediate-length dashes, short dashes, and a dotted line, respectively). No graphene coating is used and the Ag film is $34\,$nm thick. (b) Close-up of EELS peaks for the hBN substrates. The suspended sample (``vacuum substrate'') is represented by the solid line in this case. Mono-, tri-, 10-, and 20-layer hBN substrates are represented by long dashes, intermediate-length dashes, short dashes, and a dotted line, respectively. }
\end{figure}

Figure~\ref{monolayerhbnvseverything.png} depicts EELS for a $34\,$nm Ag film (with no graphene coating) on various substrates: SiO$_2$/Si, mono-layer, tri-layer, and 10-layer hBN, and semi-infinite vacuum (as would occur for a suspended sample). Introduction of the high band-gap hBN substrate dramatically enhances surface modes as compared to SiO$_2$/Si. The result is striking --- the sample with a monolayer hBN substrate has a spectral profile nearly identical to that of a suspended sample. The EELS remain very nearly identical to the suspended case for up to 10 layers. This is in accordance with experimental results that have found the electron mobility of graphene on hBN to be nearly that of suspended graphene.~\cite{Xue2011} The similarity of the hBN substrate to the vacuum is expected due to its ultra-wide bandgap, and in the case of our model partly due to the atomic layer thickness. Upon inspection of Fig.~\ref{grandhbn} (b), it is clear that both real and imaginary parts of hBN's dielectric function are small; therefore at a thickness of $3.4\,$\AA, the properties of this substrate approach those of the vacuum. The hBN substrate appears to dramatically enhance coupling to the surface plasmon resonance when compared to the SiO$_2$/Si substrate, as well as reducing the peak strength of the lossy bulk plasmon. 

\begin{figure}
\includegraphics[scale=1.0]{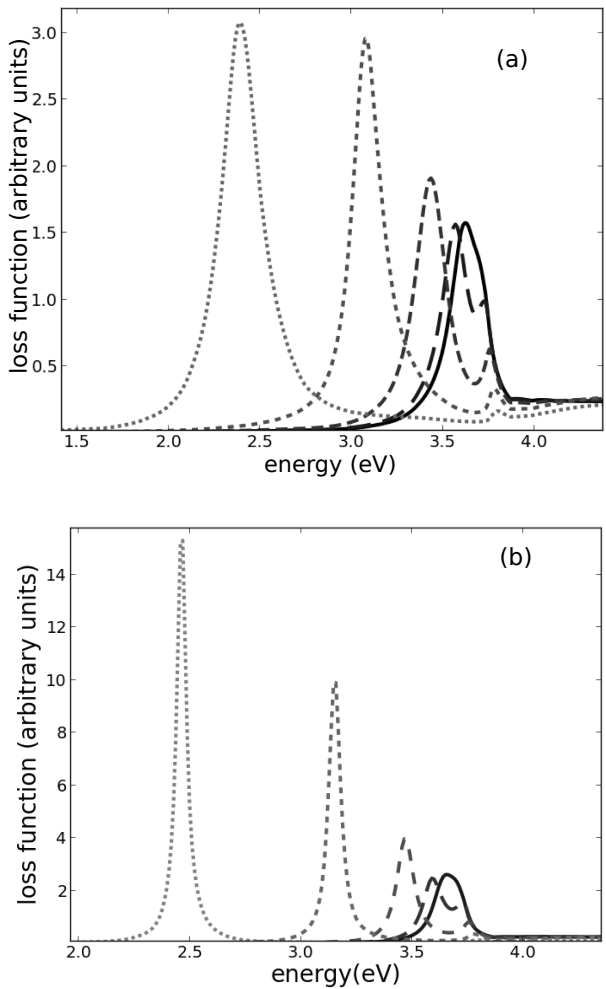}
\caption{\label{trilayergrsilverthicknesseshbn.png}(a)EELS: trilayer graphene on Ag layers (of varying thickness) with a mono-layer hBN substrate. (b) Free-standing Ag films of various thicknesses. In both (a) and (b) Ag layer thicknesses are $50\,$nm (solid line), $34\,$nm (long dashes), $20\,$nm (intermediate-length dashes), and $10\,$nm (short dashes), and $4\,$nm (dotted line).} 
\end{figure}

Fig.~\ref{trilayergrsilverthicknesseshbn.png} (a) demonstrates the combined effects of graphene coating and hBN substrate for various Ag film thicknesses. Comparison of Fig.~\ref{trilayergrsilverthicknesseshbn.png} (a) with Fig.~\ref{Sinogrsilverthicknesses} (a) reveals that the introduction of a tri-layer graphene coating and hBN substrate to the Ag film results in a surface plasmon intensity and relative surface to bulk intensity similar to a significantly thinner Ag film on the SiO$_2$/Si substrate. For example, the surface plasmon intensity and relative surface to bulk intensity of the $34\,$nm Ag case in Fig.~\ref{trilayergrsilverthicknesseshbn.png} (a) is comparable to the $20\,$nm case in Fig.~\ref{Sinogrsilverthicknesses} (a). Additionally, the surface plasmon intensity and relative surface to bulk intensity of the $20\,$nm Ag case in Fig.~\ref{trilayergrsilverthicknesseshbn.png} (a) is comparable to the $10\,$nm case in Fig.~\ref{Sinogrsilverthicknesses}(a). 
 
Comparison of Fig.~\ref{trilayergrsilverthicknesseshbn.png} (a) with Fig.~\ref{trilayergrsilverthicknesseshbn.png} (b) (which shows spectra for free-standing Ag of various thicknesses) further elucidates the effect of a few graphene layers.  For thicker Ag films, a few-layer graphene coating diminishes the bulk peak without significant broadening of the SPR.  However, for very thin Ag films, of 4nm, for example, the effect of diminishing the bulk peak is negligible and the addition of the graphene coating of three layers only slightly broadens the SPR. This is in keeping with the idea that for bulk Ag, the addition of a graphene layer enhances surface coupling, through the begrenzung effect. This is further demonstrated by the lack of significant broadening of the SPR peak for a single-layer graphene coating (see Fig. ~\ref{silvergr2} (c)).

\subsection{Au and Cu Noble-Metal Film Composites} 
\subsubsection{Spectral changes of Au and Cu films due to the graphene coating}
One may wish to employ a noble metal other than Ag in multilayer structures such as depicted in this paper. Cu has the advantage of lower cost and is therefore attractive for industrial applications. Both Cu and Au may also be of interest due to inherent surface plasmons that occur at longer wavelengths than those of Ag. In this section we demonstrate the effect of graphene coatings and the hBN substrate on Au and Cu films. For reference, Fig.~\ref{Sinogrsilverthicknesses} (b) and (c) demonstrate the effect of reduction of noble-metal film thickness for Au and Cu films (respectively) on a SiO$_2$/Si substrate. The begrenzung effect is again increasingly apparent in both Au and Cu film plasmonic response as film thickness decreases. At $20\,$nm, both Au and Cu display strong SPR and broad absorption for higher energies, rather than a bulk plasmon peak. 

\begin{figure}
\includegraphics[scale=1.0]{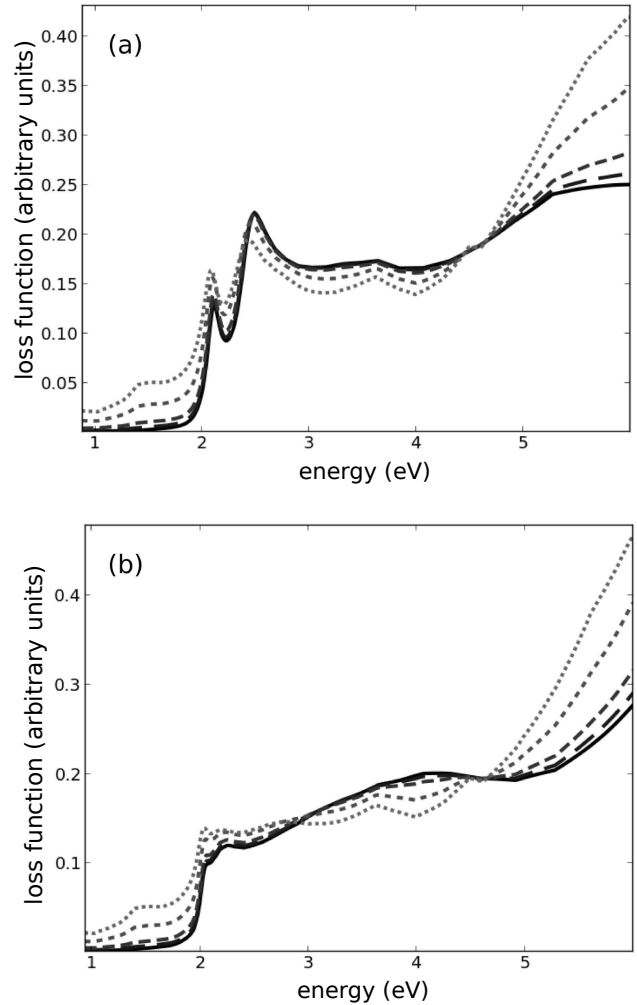}
\caption{\label{goldgraphenethickness.png}EELS: (a) The effect of various numbers of layers for the graphene film, in the case of a $34\,$nm Au layer and SiO$_2$/Si substrate, and (b) The effect of varyious numbers of layers for the graphene film, in the case of a $34\,$nm Cu layer and SiO$_2$/Si substrate. Graphene layer numbers are 0 (solid line), 1 (long dashes), 3 (intermediate-length dashes), 10 (short dashes), and 20 (dotted line).}
\end{figure}

Inspection of Fig.~\ref{goldgraphenethickness.png} (a) and (b) reveals that the introduction of a graphene coating to an Au or Cu surface does measurably strengthen surface modes while reducing bulk plasmon intensity, although the effect is not as dramatic as in the Ag case (see Fig.~\ref{silvergr2}). Coatings as thick as 20 layers further enhance and do not significantly broaden the surface peak. A jump in broad absorption (corresponding to increased losses) at higher energies is apparent in the case of both Au and Cu for more than three graphene layers. 

\subsubsection{Metallic films with hBN substrate: noble-metal comparison}
\begin{figure}
\includegraphics[scale=1.0]{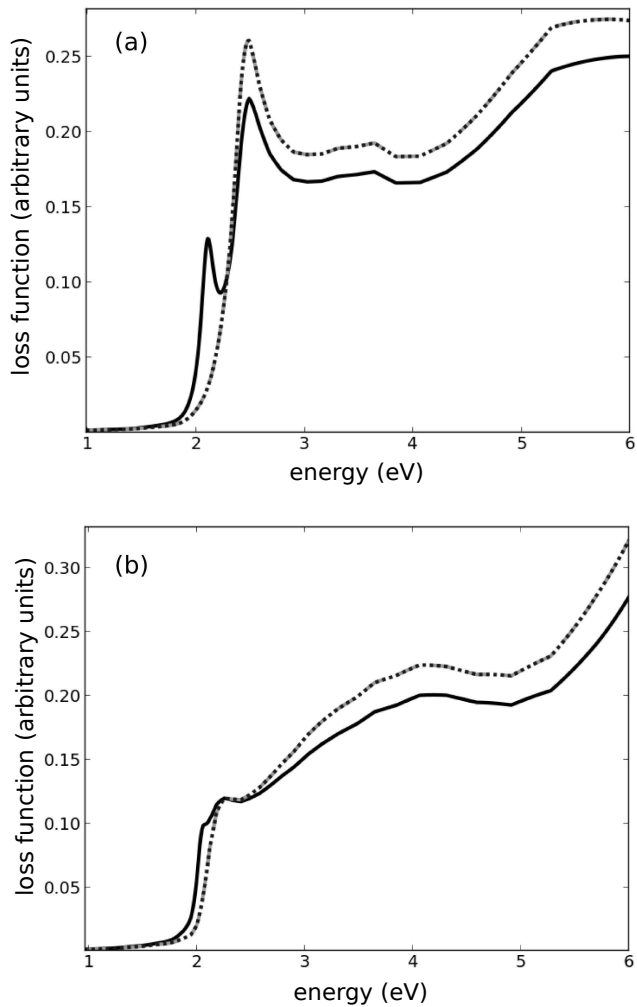}
\caption{\label{aucoatings.png}EELS: $34\,$nm Au (a) and Cu (b) films, with monolayer graphene coatings and various substrates including SiO$_2$/Si (solid line), hBN monolayer (dashes), and vacuum (dotted line) substrates.}
\end{figure}
Figure~\ref{aucoatings.png} (a) and (b) are the EELS for Au and Cu films of thickness $34\,$nm (respectively), with monolayer graphene coating, comparing the SiO$_2$/Si substrate, hBN substrate, and suspended samples. Surface peaks at $\approx 2\,$eV for both Au and Cu are surface peaks due to change in dielectric function across the metal/SiO$_2$/Si substrate interface. The addition of the hBN layer in these cases appears to sharpen and enhance Au and Cu inherent plasmons. Notably, in the Au and Cu cases, the hBN substrate appears to produce a similar spectral profile to the suspended ``vacuum substrate'', as in the case of the Ag film. Rigorous optimization of both the surface coating and substrate for both Au and Cu films is the subject of further investigation. 

\section{DISCUSSION}
In this study we systematically investigated the effect of the use of different plasmonic materials, different semiconducting substrates, and different layer thicknesses in a multilayer graphene-based plasmonic composite structure. For mono-, bi-, and tri-layer graphene, in the case of $34\,$nm Ag layers (where bulk modes would dominate in the absence of a graphene coating), bulk plasmon modes are significantly diminished, while maintaining the strength of surface modes. 

Reduced plasmon losses for graphene coated plasmonic metals on hBN substrates likely originates within the context of our mesoscopic model from two primary physical effects: (1) The addition of a graphene boundary layer on the metallic surface reduces coupling of excitations to bulk plasmons through the begrenzung effect. The origin of the begrenzung effect is a reduction of the degrees of freedom for excitations, and thus further surface confinement comes at the expense of bulk oscillations, leading to reduced losses. (2) The strongly insulating hBN substrate diminishes bulk losses and enhances surface confinement particularly through a reduction in scattering. 

Traditional substrates such as Si are known to degrade the electron transport properties of graphene as compared to suspended samples.~\cite{Xue2011} Reduced transport properties are due to various scattering mechanisms. There is strong evidence that scattering in these systems is due in large part to various substrate interactions including interfacial phonons (which are taken into account in our work), surface charge traps, and substrate stabilized ripples.~\cite{Bolotin2008351} 

It is therefore reasonable to mitigate scattering in graphene-based systems for plasmonics (and devices for plasmonics in general) by suspending samples, or through the use of alternative substrates.~\cite{Bolotin2008351} In this work, for Ag films, the use of a hBN substrate (rather than the more traditional silicon) is found to shift plasmonic coupling towards surface modes, both diminishing bulk losses and enhancing the SPR peak amplitude. This  effect is particularly dramatic in the case of the $34\,$nm Ag film (including those with no graphene top-coating), as an Ag film at this thickness on an SiO$_2$/Si substrate produces EELS that are extremely dominated by the bulk resonance.~\cite{Johnson1972} This is interesting in light of recent experimental results, showing electron mobility of graphene on hBN to be similar to that of suspended graphene.~\cite{Xue2011} We expect SPR enhancement for other wide-bandgap substrates due to enhanced surface plasmon field confinement. Further comparison of alternate wide-bandgap substrate materials, such as SiC, are the subject of future work.

As expected, bulk modes were quenched for $4\,$nm Au and Cu films. For Ag, bulk modes very nearly vanished at this thickness, and are nearly undetectable when coupled with the hBN substrate. However, the morphology of very thin metallic films deposited on graphene is difficult to control, often forming island-like structures of various sizes.~\cite{Boragno2009} This is also the case for graphene deposited on a metallic substrate, as agglomeration below a critical thickness is a general property of thin films.~\cite{Rha1997} Results of this study indicate that Ag films as thick as $34\,$nm, when coated with a graphene film and placed on a wide-bandgap substrate such as hBN, may also be employed for low-loss plasmon resonance applications. An hBN substrate, due to its ultra-flat morphology, would also lend additional uniformity to the structure. 

The mesoscopic model used in these calculations has several limitations that are worth discussing. Results of this study are valid in the long-wavelength limit for which the continued fraction expression by Lambin et al. was derived. Additionally, coupling between layers is classical (via boundary conditions), and as a result inter-layer hopping is neglected. This tunnelling has been determined by angle-resolved photoemission spectroscopy to be important for graphene band-structure, leading to $\pi$ band splitting, which increases with layer number ($\approx 0.7\,$eV at 4 layers).~\cite{interlayer2} Multilayer graphene excitation spectra have been calculated by Ohta et al. with the inclusion of inter-layer hopping in the kinetic Hamiltonian. Tunneling is found to be primarily important in accurately producing the low energy region of the excitation spectrum ($\omega\sim v_{F}k$), where $v_{F}$ is the Fermi velocity).~\cite{interlayer1} Though tunnelling is not as critical in the regime of interest for this work, calculations for graphene-based multilayer plasmonic systems geared towards lower-energy applications should take care to include inter-layer hopping.

More general EELS, accurate for a wider range of $k$ values and angles of incidence will require first principles calculations for the entire structure, and incorporate higher level quasiparticle interactions (including the effect of excitons). Using guidance from the results of this study, future work consisting of 1-3 layers of graphene on silver, with the hBN substrate are the subject of future work. The band-gap for hBN is underestimated by about 33\% in the LDA~\cite{PhysRevB.76.073103} --- the effect of hBN on surface confinement of plasmons may be even more dramatic than is demonstrated in this work. A quasiparticle GW correction to the LDA calculation will be employed, as it has been demonstrated to bring the hBN bandgap into close agreement with experimental results.~\cite{PhysRevLett.96.026402} 

As the individual layers in these structures are nearly isomorphic, rather than exactly isomorphic, future work will incorporate the effect of lattice strain on the optical properties of the composite. Strain engineering is expected to provide a further means of plasmon resonance tunability.~\cite{ciammarella2010} Additionally the effect of configurations other than the flat-armchair morphology for graphene is the subject of ongoing work. The effect of rippling and defects in graphene-based multilayer structures is of particular interest, as is a rigorous quantification of the uncertainties involved in the modelling and manufacturing stages, and optimal design against those uncertainties.

\begin{acknowledgments}
The authors would like to thank Katie Rice, Ann Chiaramonti Debay, and Alex Smolyanitsky for helpful discussions. This research was performed while the first author held a National Research Council Research Associateship Award at the National Institute of Standards and Technology. This work represents an official contribution of the National Institute of Standards and Technology and is not subject to copyright in the USA. 
\end{acknowledgments}

\providecommand{\noopsort}[1]{}\providecommand{\singleletter}[1]{#1}%

\end{document}